\documentclass[prd,aps,floatfix,notitlepage,prl,reprint,superscriptaddress,twocolumn,nofootinbib]{revtex4-1}
\usepackage{graphicx,graphics,amsmath,bm,natbib,multirow,hyperref,float,fancyhdr,amssymb}
\usepackage[usenames,dvipsnames]{color}
\usepackage[caption = false]{subfig}
\hypersetup{colorlinks=true,citecolor=blue,linkcolor=blue}

\newcommand{\aj}{{\em Astron.\ J.}}
\newcommand{\aap}{{\em Astron.\ Astrophys.}}
\newcommand{\apjl}{{\em Astrophys.\ J.\ Lett.}}
\newcommand{\apjs}{{\em Astrophys.\ J.\ Suppl.}}
\newcommand{\mnras}{{\em Mon.\ Not.\ R.\ Astron.\ Soc.}}

\newcommand{\physrep}{{\em Phys.\ Rep.}}
\newcommand{\pasj}{{\em Publ.\ Astron.\ Soc.\ Japan}}
\newcommand{\jcap}{{\em J.\ Cosmol.\ Astropart.\ Phys.}}

\newcommand{\be}{\begin{equation}}
\newcommand{\ee}{\end{equation}}

\begin{document}

\title{Detection of anisotropic galaxy assembly bias in BOSS DR12} 
\author{Andrej Obuljen}
\affiliation{Waterloo Centre for Astrophysics, University of Waterloo, 200 University Ave W, Waterloo, ON N2L 3G1, Canada}
\affiliation{Department of Physics and Astronomy, University of Waterloo, 200 University Ave W, Waterloo, ON N2L 3G1, Canada}
\author{Will J. Percival}
\affiliation{Waterloo Centre for Astrophysics, University of Waterloo, 200 University Ave W, Waterloo, ON N2L 3G1, Canada}
\affiliation{Department of Physics and Astronomy, University of Waterloo, 200 University Ave W, Waterloo, ON N2L 3G1, Canada}
\affiliation{Perimeter Institute for Theoretical Physics, 31 Caroline St. North, Waterloo, ON N2L 2Y5, Canada}
\author{Neal Dalal}
\affiliation{Perimeter Institute for Theoretical Physics, 31 Caroline St. North, Waterloo, ON N2L 2Y5, Canada}

\begin{abstract}

We present evidence of anisotropic galaxy assembly bias in the Baryon Oscillation Spectroscopic Survey Data Release 12 galaxy sample at a level exceeding $5\sigma$. We use measurements of the  line-of-sight velocity dispersion $\sigma_\star$ and stellar mass $M_\star$ to perform a simple split into subsamples of galaxies. We show that the amplitude of the monopole and quadrupole moments of the power spectrum depend differently on $\sigma_\star$ and $M_\star$, allowing us to split the galaxy sample into subsets with matching monopoles but significantly different quadrupoles on all scales. Combining data from  the LOWZ and CMASS NGC galaxy samples, we find $>5\sigma$ evidence for anisotropic bias on scales $k<0.15\,h\,\rm{Mpc}^{-1}$.  We also examine splits using other observed properties. For galaxy samples split using $M_\star$ and projected size $R_0$, we find no significant evidence of anisotropic bias. Galaxy samples selected using additional properties exhibit strongly varying degrees of anisotropic assembly bias, depending on which combination of properties is used to split into subsets.  This may explain why previous searches for this effect using the Fundamental Plane found inconsistent results. We conclude that any selection of a galaxy sample that depends on $\sigma_\star$ can give biased and incorrect Redshift Space Distortion measurements.

\end{abstract}

\maketitle

\section{Introduction}

The main goal of current and future galaxy redshift surveys is to extract cosmological information from the observed galaxy density field. This information is encoded in the underlying dark matter density field, the galaxy bias and the projection from comoving positions to observed quantities -- galaxy angular positions and redshifts. 

Galaxy bias is the term used to describe the connection between galaxies and dark matter. On sufficiently large scales, galaxy clustering is linearly biased compared to matter clustering \cite{Kaiser_bias,Bias_review,Wechsler_review}. The linear galaxy bias is often assumed to be a function of only host halo mass and redshift  \cite{Press1974,Bond1991,Mo1996} --- halos of fixed mass are more clustered at higher redshifts, while at fixed redshift, more massive halos are more strongly clustered than less massive halos \cite{Kaiser_bias}.
Biases of this form may be expressed, on linear scales, using the lowest-order expansion
\begin{equation}
    \delta_g \approx b\,\delta_m,
\end{equation}
where $\delta_m$ is the matter overdensity, $\delta_g$ is the galaxy (or halo) overdensity, and $b$ is the linear bias, which tends towards constant behavior on large scales.  Written this way, we can see that statistical homogeneity and isotropy of the matter field ensure statistical homogeneity and isotropy of the galaxy field.

However, because we observe galaxies in redshift-space, and not in real space, the observed clustering of galaxies can violate statistical isotropy, an effect called redshift-space distortion (RSD).  
RSDs result in an anisotropic galaxy power spectrum, with power boosted in the radial (line-of-sight) direction compared to the transverse direction on linear scales according to  \cite{Kaiser}
\begin{equation} 
        \delta_g \approx (b + f\,\mu^2)\delta_m,
\label{kaiser_rsd}
\end{equation}
where $f$ is the growth rate and $\mu = k_\parallel/k$.
Measuring the amplitude of the anisotropy  provides a way to measure the growth rate as a function of redshift using large-scales, commonly parameterised by $f\sigma_8$. These RSD measurements are then contrasted with predictions of general relativity, and are thus useful to test gravity on large-scales (e.g.\ \cite{2008Natur.451..541G}).

With galaxy surveys probing increasingly larger volumes, we are entering a regime where it is necessary to take into account or mitigate all known systematic effects in order to both precisely and accurately extract key cosmological parameters from the observed 3D galaxy distribution. Numerical simulations have now established that halo bias correlates with halo properties, e.g.\ history and formation time, concentration, spin, large-scale tidal field, etc., an effect termed assembly bias  \cite{Gao2005,Gao2007,Dalal2008,Hahn2009}. 
Selections of halos based on scalar or tensor halo properties lead to different assembly bias effects. Selections on scalar properties that are independent of halo orientation change the bias in Eq.~\eqref{kaiser_rsd}, but do not change the form of this equation.
However, the assumption that the galaxy bias $b$ in Eq.~\eqref{kaiser_rsd} is a scalar that is independent of the direction of the wavevector $\bm{k}$, no longer necessarily holds for a sample of halos selected on their orientation and therefore, for example, on the tidal field. In this paper, we focus on measuring this anisotropic assembly bias (AB) signal, as is potentially caused by large-scale tidal fields.

AB poses a potential problem for RSD measurements, as first discussed in \cite{Hirata_th}, through the correlation of galaxy (non-scalar) properties with large-scale tidal fields. The AB signal, as discussed further in \S\ref{sec:Theory}, is degenerate with the RSD signal, and selection effects present in galaxy samples can therefore act as a contaminant for RSD-based growth factor measurements from the clustering.

This effect has the same root cause as intrinsic alignments (IA), a contaminant of weak lensing measurements (see review by \citealt{Troxel2015}). Due to IA, galaxy ellipticities are correlated both between pairs of galaxies (II correlations) and between galaxies and the large-scale tidal field (GI correlations; \cite{Hirata2004}). Locally, IA therefore mimic the weak lensing signal, but they have different scaling with galaxy redshift allowing them to be separated \cite{Joachimi2008}. For weak lensing, it is only the clustering transverse to the line-of-sight that is important. In contrast, for RSD, the cosmological signal depends on the anisotropy with respect to the line-of-sight. Thus, although IA and AB have a common root cause, that of tidal fields driving galaxy properties, they manifest upon cosmological measurements in different ways.

In order to understand and mitigate the effect of AB upon RSD measurements, we would like to know the intrinsic anisotropic clustering signal for a sample of galaxies. As RSD and AB are perfectly degenerate, it is not possible to disentangle them for a given sample. However, we can measure AB by splitting a sample as a function of a tensor property of the galaxies that is expected to correlate with the anisotropic tidal field. In this case, the RSD signal is unaffected, while the sub-samples will exhibit different AB.

The Baryon Oscillation Spectroscopic Survey (BOSS) \cite{BOSS} provides the largest galaxy redshift survey obtained to date in terms of number of spectroscopic redshift measurements, and the best chance of measuring the AB signal using a differential technique. The BOSS galaxies are predominantly Luminous Red Galaxies, that are expected to be most strongly correlated with the tidal fields \cite{Chisari2015,Hilbert2017}. Previous efforts to measure the radial alignment of BOSS galaxies exist in literature \cite{Hirata_obs,Singh}. However, these studies have used the residuals from the Fundamental Plane to split the catalogue, and show inconsistencies in their final results, with a tentative ($2.3\sigma$ level) detection reported by \cite{Hirata_obs} that is not confirmed by a similar analysis by \cite{Singh}. 

In order to understand AB further using the BOSS sample, we consider different ways to use the properties of BOSS galaxies to artificially split the sample into two subsamples and compare how the large-scale monopole and quadrupole moments of the power spectrum (hereafter monopoles and quadrupoles) depend on this split. We demonstrate that the results are very sensitive to the exact form of the split, possibly explaining why previous analyses have reached different conclusions. We show that a robust split is able to measure AB at a level exceeding $5\sigma$ significance. 

In \S\ref{sec:Theory} we start by reviewing the theory underlying modelling the AB and linear power spectrum, and the mechanism by which we intend to measure AB. In \S\ref{sec:Data} we outline the BOSS data and mocks used in our analysis, including the measurements of galaxy properties that we use to split our sample. The method for making clustering measurements is described in \S\ref{sec:pk}. There are many ways to split the sample to try to measure AB, and we have found that the strongest measurements arise when using the radial velocity dispersion - results from splits based on this measurement are presented in \S\ref{sec:Matching_P0}, with some technical details left for Appendix~\ref{sec:Cuts}. In Appendix~\ref{sec:Matching_P2}, we discuss alternative splits based on observed size $R_0$ and the Fundamental Plane. We discuss our results in \S\ref{sec:conc}.

Throughout this paper we assume the following flat $\Lambda\mathrm{CDM}$ cosmology: $h=0.676$, $\Omega_\mathrm{m}=0.31$, $\Omega_\mathrm{b}=0.048$, $\Omega_\Lambda=0.69$, $n_s=0.9667$, $\sigma_8=0.834$ and $T_\mathrm{cmb}=2.73\,\mathrm{K}$, in agreement with the results from Planck \cite{planck15}.

\section{Theory}
\label{sec:Theory}

\subsection{Linear bias model}

At linear order on subhorizon scales we assume that the observed galaxy overdensity field in Fourier space may be related to the matter overdensity field $\delta_m$ by \cite{Bias_review}
\begin{equation}
\label{eq:linearbias}
\delta_g(\bm{k}) = (b_g + f\mu^2)\delta_m(\bm{k}) + b_{ij} s_{ij}(\bm{k}),
\end{equation}
where $b_g$ is the linear scalar galaxy bias, $b_{ij}$ are linear anisotropic bias coefficients, $f=d \ln D(a)/d \ln a$ is the logarithmic growth rate, $D(a)$ is the linear growth factor, $\mu=k_\parallel/k$, $b_{ij}$ represents the galaxy AB, and $s_{ij}(\bm{k})=(k_i k_j/k^2 - \delta_{ij}/3)\delta_m(\bm{k})$ is the traceless  tidal tensor.  If our galaxy sample is independent of the transverse properties of galaxies (e.g., independent of projected galaxy ellipticity on the sky), then we can neglect all components of $b_{ij}$ except the parallel component, which we write as $b_q$, simplifying Eqn.\ \eqref{eq:linearbias} to
\begin{equation}
\label{eq:delta}
\delta_g(\bm{k}) 
=\left(b_g + f\mu^2 + b_q (\mu^2-1/3)\right)\delta_m(\bm{k}).
\end{equation}
The first terms in Eqn.~\eqref{eq:delta} are the combination of standard linear galaxy bias and the Kaiser redshift-space distortions from Eqn.\ \eqref{kaiser_rsd} \cite{Kaiser_bias,Kaiser,Bias_review}. The final term is related to the orientation-dependent selection effects in the presence of shape correlations with the large scale tidal field, and is usually assumed to vanish. This is justified under the assumption that non-scalar properties of galaxies (e.g.\ shapes, velocity dispersion or angular momenta) are uncorrelated with the large-scale tidal field. Even in the presence of the non-zero correlations with the tidal field, this term vanishes if the sample is complete in all shape orientations. However, both of these assumptions may not necessarily be justified in observations. Previous studies used numerical simulations to show that halos shape, velocity dispersion and spin have non-zero correlations with the tidal field \cite{Obuljen}. Provided that galaxy non-scalar properties correlate with those of their host halos, orientation-dependent galaxy selection effects could make the $b_q$ term in Eqn.~\eqref{eq:delta} non-zero. 

The level of completeness in orientation-dependent galaxy selection is difficult to ascertain. One way to examine the completeness would be to use hydro-dynamical simulations that provide the galaxy shapes and orientations, and apply the same target selection algorithm used for the observed targets. However, different simulations give different galaxy-halo orientation correlations \cite{Bryan2013,Tenneti2014,Tenneti2016}. Furthermore, the issue of completeness is more important for lower mass galaxies which are near the detection threshold and are more numerous.

The net effect of the orientation-dependent selection effects on the power spectrum at the linear level is indistinguishable from the RSD effect for any single sample of galaxies. Using Eqn.\ \eqref{eq:delta}, the galaxy power spectrum takes the following form on linear scales:
\be \label{eq:P2D}
P_g(k,\mu) = \left(b_g -\frac{b_q}{3} + (f+b_q)\mu^2\right)^2 P_m(k),
\ee
where $P_m$ is the matter power spectrum. 

\subsection{Model independent analysis}

There are a number of methods that could be used to compare multipoles from two different samples. One would be to fit models like Eqn.\ \eqref{eq:P2D} to the measurements, allowing both $b_q$ and $b_g$ to vary. This would allow us to compare samples with different window functions (angular and radial distributions). In the fits, the cosmological term $f$ in Eqn.\ \eqref{eq:P2D} will be the same for different samples of galaxies, while $b_q$ and $b_g$ will vary, and we could test whether $b_q\ne0$ is required for one of the samples.

Instead, if we construct samples with the same angular and redshift distribution (so they have matching window functions), then we can choose to be agnostic about the particular model of RSD to use. In the absence of anisotropic bias, two samples with the same window function and identical monopoles would be expected to also have matching quadrupoles. One way to see this is by noting that velocity bias is negligible on large scales: numerical simulations show that any halo velocity bias at $k<0.2\,h\,{\rm Mpc}^{-1}$ is $\lesssim 1\%$ at $z<1$ \cite{Velocity_bias}, as expected from the equivalence principle in general relativity.  Therefore, all objects have the same bulk velocities on large scales, which means that all objects transform in the same way between real space and redshift-space, on linear scales.  Therefore, a population's redshift-space power spectrum is determined solely by its real-space power spectrum.  In the absence of anisotropic bias, statistical isotropy of the matter field ensures that the power spectrum of the galaxy field is isotropic in real space, which means that two samples with matching real-space power spectra will have matching redshift-space multipoles for all $\ell$. 

Given that tests of the match of two measurements are more robust than model fitting, this is our preferred method for detecting the effect of AB on data. That is, we construct samples with matching windows and monopoles and simply test whether the quadrupoles also match. Any mismatch would be a signal for AB. We could have equivalently chosen to construct samples with matching quadrupoles, and compare their monopoles. 
\section{Data}
\label{sec:Data}
In this section we describe the galaxy catalog, galaxy property measurements and the galaxy mock catalogs used in our analysis.

\subsection{BOSS DR12 sample}  \label{sec:BOSS}

The Baryon Oscillations Spectroscopic Survey (BOSS) \cite{BOSS}, part of SDSS-III \cite{SDSSIII}, measured spectroscopic redshifts for over a million of galaxies in the redshift range $0.15<z<0.7$. We use the publicly available DR12 release\footnote{\href{https://data.sdss.org/sas/dr12/boss/lss/}{https://data.sdss.org/sas/dr12/boss/lss/}} \cite{Alam_2015} which contains the LOWZ and CMASS galaxy samples. These galaxy samples were obtained using two different targeting algorithms based on color-cuts and flux limits \cite{2016MNRAS.455.1553R}. LOWZ sample selection contains bright, red galaxies at lower redshift $0.15<z<0.43$. CMASS was designed to be a stellar-mass limited sample covering the redshift range $0.43<z<0.70$. The majority of the CMASS sample are central, red, elliptical galaxies hosted in halos of masses $\sim 10^{13}\,M_\odot/h$ \cite{2011ApJ...728..126W}.

We use data and random LSS catalogs of both CMASS and LOWZ samples. Each sample is further split into North and South Galactic Cap (NGC and SGC). We limit our analysis to the NGC parts of the CMASS and LOWZ samples, which contain 568776 and 248237 galaxies, respectively. Due to the smaller sky coverage, the SGC samples are equivalent to $37\%$ and $46\%$ of the CMASS and LOWZ NGC samples, respectively. Since we measure the clustering properties of selected subsamples in Fourier space where the window function is important, including the SGC would lead to more structure within the window, and complicate the interpretation of our results. Additionally, there are fundamental differences between the SGC and NGC galaxy samples: they have different bias and densities of legacy targets (those not imaged by BOSS because of a prior secure redshift measurement). In order to make our results as robust as possible, we do not include the SGC galaxies in our analysis.

Each BOSS galaxy is assigned three incompleteness weights to account for different systematic effects present in the dataset \cite{Ross_2012,Anderson_2014,2016MNRAS.455.1553R}. We account for these in our power spectra measurements by using the following completeness weight for each galaxy:
\be
w_c=w_\mathrm{sys}(w_\mathrm{rf}+w_\mathrm{fc}-1),
\ee
where $w_\mathrm{sys}$ is the systematic weight accounting for the seeing condition and stellar weight, $w_\mathrm{rf}$ is the redshift failure weight and $w_\mathrm{fc}$ is the fiber collision weight.

\begin{figure}[!ht]
\includegraphics[width=0.48\textwidth]{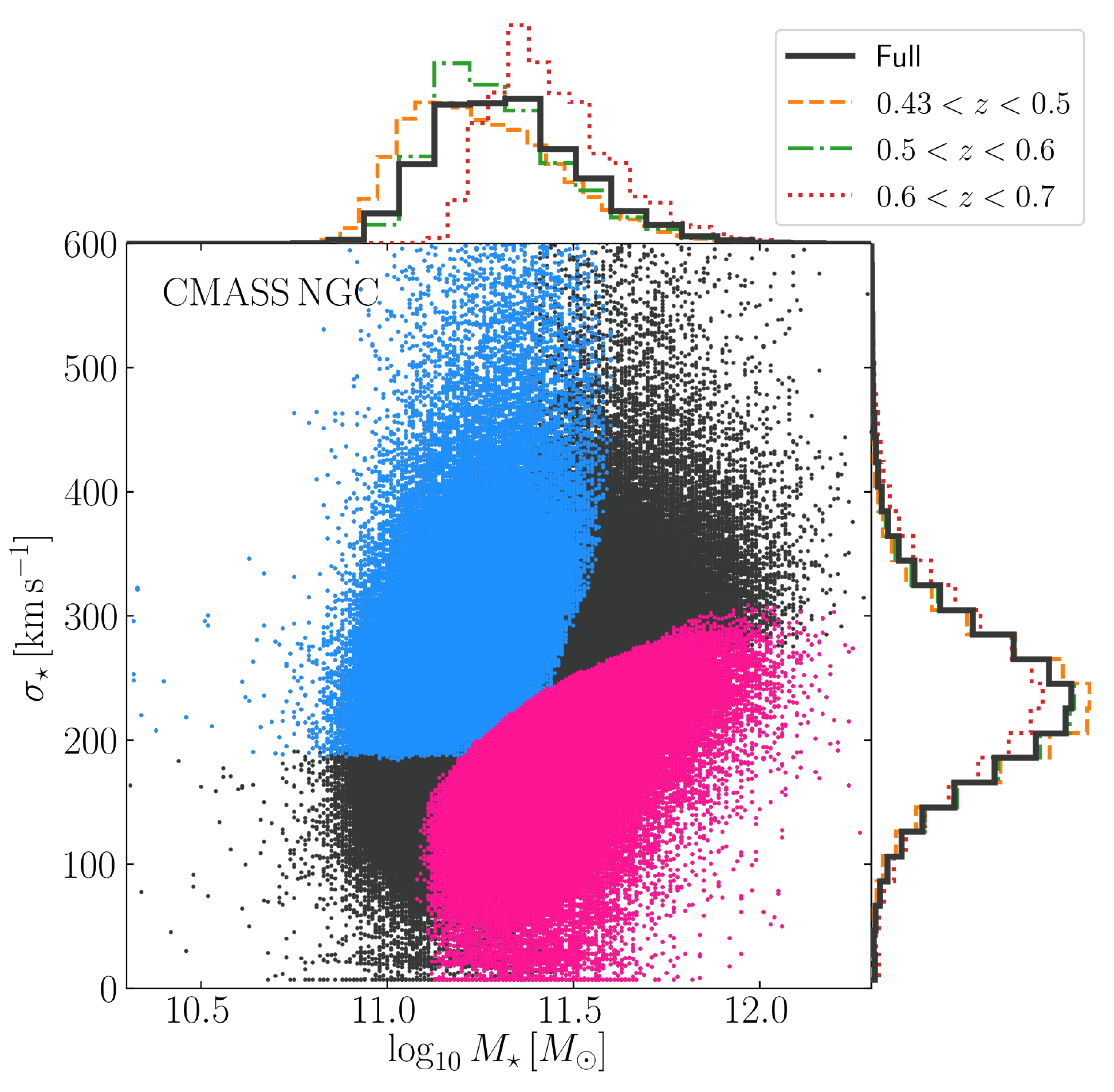}
\caption{Stellar velocity dispersion versus stellar mass for the galaxies in the CMASS NGC sample (central panel). The histograms on the top and right show the distributions of stellar mass and velocity dispersion, respectively, for the full sample (solid blue line) and in several redshift bins (other lines).  In the scatterplot, point colors indicate which subset each galaxy was assigned to (see \S\ref{sec:Matching_P0}), either light blue (high $\sigma_\star$, low $M_\star$), pink (low $\sigma_\star$, high $M_\star$), or black (the rest).  Although the subsamples appear to overlap in this projection, that is an artifact of the redshift dependence of the $M_\star$ and $\sigma_\star$ distributions.  In the 3D space of ($M_\star$, $\sigma_\star$, $z$), the subsets are well separated, as depicted in Fig.\ \ref{fig:Pixel}. }
\label{fig:Mstar_sigmastar}
\end{figure}

\subsection{Galaxy properties} \label{sec:galprop}
BOSS galaxies' properties were also measured using the observed spectra. We use the Portsmouth extended galaxy catalog\footnote{\href{https://www.sdss.org/dr12/spectro/galaxy_portsmouth/}{www.sdss.org/dr12/spectro/galaxy\_portsmouth/}} which contains the measurements of galaxy velocity dispersion \cite{2013MNRAS.431.1383T} and stellar masses \cite{2013MNRAS.435.2764M}. For the stellar masses, we use the measurements obtained using the passive model with the Kroupa initial mass function. We match these extended catalogs to the main catalog using the following columns: FIBERID, PLATE and MJD. We find most of the galaxies ($>99.9\%$) in the LSS catalogs to have matching measurements of the stellar properties. Instead of removing the galaxies that do not have the matching stellar properties from further analysis, we randomly sample the missing values from the known distribution of $\sigma_\star$ and $M_\star$. We do this in order not to change the sky coverage, albeit at the expense of making any anisotropic signal we are after slightly weaker.

In Fig.\ \ref{fig:Mstar_sigmastar} we show the velocity dispersion and stellar mass distributions for the CMASS NGC galaxies. While the velocity dispersion measurements show weak dependence on redshift, inferred stellar masses exhibit stronger redshift evolution.

The majority of the spectra associated with the galaxies in the samples were obtained using the BOSS spectrograph. However, about a third of LOWZ sample were legacy objects with spectra obtained using the previous SDSS I/II spectrograph. The main difference is the angular radius of the fibre --- $r_\mathrm{fiber}=1"$ in the case of BOSS and $r_\mathrm{fiber}=1.5"$ in the case of SDSS I/II. A fixed fiber size covers different parts of the galaxy velocity dispersion profile at different distances and an aperture correction (AC) is usually applied to the velocity dispersion measurement \cite{1995MNRAS.276.1341J,1999MNRAS.305..259W,Singh}:
\be
\label{eq:sigma_star_AC}
\sigma_{\star}^\mathrm{AC} = \sigma_{\star}\left(\frac{r_\mathrm{fiber}}{r_\mathrm{cor}/8}\right)^{0.04},
\ee
where $r_\mathrm{cor}=r_\mathrm{deV}\sqrt{q_{b/a}}$, $r_0$ is the effective radii (in arcseconds) and $q_{b/a}$ is the axis ratio obtained from best fit models. For both $r_\mathrm{deV}$ and $q_{b/a}$ we use the results from the de Vaucouleurs model fits in the $i$-band to measure the amplitude of these adjustments. This typically results in 10\% higher values of $\sigma_\star$ and for the CMASS sample it is not expected to affect our results given the low number of legacy targets. We therefore do not include these corrections in our baseline results, although we do test the impact of this correction on our LOWZ results.

We also investigate the clustering dependence in terms of the projected galaxy sizes. We compute the galaxy physical radius using $r_\mathrm{cor}$ as $R_0=D_A(z)\tan(r_\mathrm{cor})\times 10^3$, where $D_A$ is the angular diameter distance. 

\subsection{Mock galaxy catalogs} \label{mocks}
We use the BOSS-LRG DR12 MultiDark-Patchy 
mock galaxy catalogs\footnote{\href{http://www.skiesanduniverses.org/page/page-3/page-15/page-9/}{www.skiesanduniverses.org/page/page-3/page-15/page-9/}}\cite{Kitaura, Rodriguez-Torres}. These catalogs were produced to match the spatial distribution and the clustering properties of observed galaxies. We make use of these catalogs to compute the covariance matrices used in our analysis.

\section{Power spectrum measurements}  \label{sec:pk}
We measure the multipoles of the auto and cross power spectrum using the FFT-based algorithm from \cite{Hand} as implemented in \texttt{nbodykit} \cite{nbodykit}. This algorithm builds upon previous estimators \cite{Bianchi,Scoccimarro} and allows for the fast evaluation of the power spectrum estimator from \cite{Yamamoto}. We briefly describe the algorithm we use here. 

The weighted galaxy density field is defined as \cite{FKP}:
\be
F(\bm{r})=\frac{w(\bm{r})}{I^{1/2}}[n(\bm{r}) - \alpha n_s(\bm{r})],
\ee
where $w(\bm{r})$ is the general weighting scheme, $n$ and $n_s$ are the number density of observed galaxies and the synthetic random catalog, respectively, $\alpha$ is the ratio of number of observed galaxies to the total number of objects in the random catalog, while the normalization factor is given by $I\equiv\int d\bm{r}[w(\bm{r})n(\bm{r})]^2$. We use the following total weights for both the data and random:
\be
w=w_c\times w_\mathrm{FKP},
\ee
where $w_\mathrm{FKP}(z)\equiv(1+n(z)P_0)^{-1}$ are the standard FKP weights \cite{FKP} and we adopt $P_0=10^4\,h^{-3}\, \mathrm{Mpc}^3$.

The power spectrum estimator in \cite{Hand} for the multipole $\ell$ is defined as:
\be
\widehat{P}_\ell(k) = \frac{2\ell+1}{I}\int \frac{d\Omega_k}{4\pi}F_0(\bm{k})F_\ell(-\bm{k}),
\ee
where:
\be
\begin{split}
F_\ell(\bm{k}) & \equiv \int d\bm{r} \ F(\bm{r}) e^{i \bm{k} \cdot \bm{r}} \mathcal{L}_\ell(\hat{\bm{k}} \cdot \hat{\bm{r}}), \\ & = \frac{4\pi}{2\ell+1} \sum_{m=-\ell}^{\ell} Y_{\ell m}(\hat{\bm{k}}) \int d\bm{r} F(\bm{r}) Y_{\ell m}^*(\hat{\bm{r}}) e^{i \bm{k} \cdot \bm{r}},
\end{split}
\ee
where $\Omega_k$ is the solid angle in Fourier space, $\mathcal{L}_\ell$ is the Legendre polynomial of order $\ell$ and $Y_{\ell m}$ are spherical harmonics. 

When computing the auto power spectrum of galaxy subsamples, we use the FKP weights computed using the subsample's radial distribution $n_\mathrm{sub}(z)$, for both the data and the random catalogs. Furthermore, we make use of the existing parent random catalogs \cite{2016MNRAS.455.1553R}, in which we keep the total number and the angular distribution of objects, while we obtain the radial distribution by randomly sampling redshifts from the galaxy subsample. We do this in order to match the shape of the random radial distribution to that of the subsample, whilst not changing the distribution on the sky which contains information on the survey mask \cite{2016MNRAS.455.1553R}. The radial integral constraint required to correct this procedure will be the same for both subsamples, and therefore does not affect our comparison of the two quadrupoles \cite{demattia2019}.

When computing the cross power spectrum of a galaxy subsample with its parent catalog, we assign the same parent data and random catalog to two different meshes. One mesh contains all the objects from the parent data and random catalogs with the corresponding weights. In the second mesh, we use non-zero weights only for the data in the subsample and use the FKP weights computed with $n_\mathrm{sub}(z)$. For the random catalog, we uniformly sample from the parent random catalog a fraction of objects matching the fraction of galaxies in the subsample, compared to the total number of galaxies and give zero weights to the rest of the objects. We use $n_\mathrm{sub}(z)$ to assign the FKP weights to the random catalog. Finally, we cross-correlate the two fields to obtain the cross power spectrum.

All power spectrum measurements were made with $k_\mathrm{min} =0.01\,h\,\rm{Mpc}^{-1}$ using linearly spaced bins with the bin size $\Delta k =0.01\,h\,\rm{Mpc}^{-1}$. We use triangular-shaped cloud interpolation \cite{HockneyEastwood} to assign galaxies to a mesh of $512^3$ cells, de-convolve the effects of interpolation scheme on the measured power spectrum \cite{2005ApJ...620..559J} and make use of interlacing technique to reduce the effects of aliasing \cite{HockneyEastwood,Sefusatti}. We restrict our analysis to the monopole and quadrupole multipoles. Furthermore, we limit the largest scales we use for fitting following the previous analysis of the same datasets \cite{BOSS_RSD}, which was based on the impact of systematic weights. For the monopoles we use $k>0.02\,h\,\mathrm{Mpc}^{-1}$ and for the quadrupoles we use $k>0.04\,h\,\mathrm{Mpc}^{-1}$ \cite{BOSS_RSD}.

The constant shot noise term is computed following \cite{Beutler14} and accounts for the completeness and FKP weights. We subtract this term from all of our monopole measurements. Note that in our cross power spectrum measurements of the subsamples with the full sample, the shot noise is expected to be similar to the shot noise of the full sample \cite{Peebles}.

\begin{figure*}[!ht]
\includegraphics[width=0.96\textwidth]{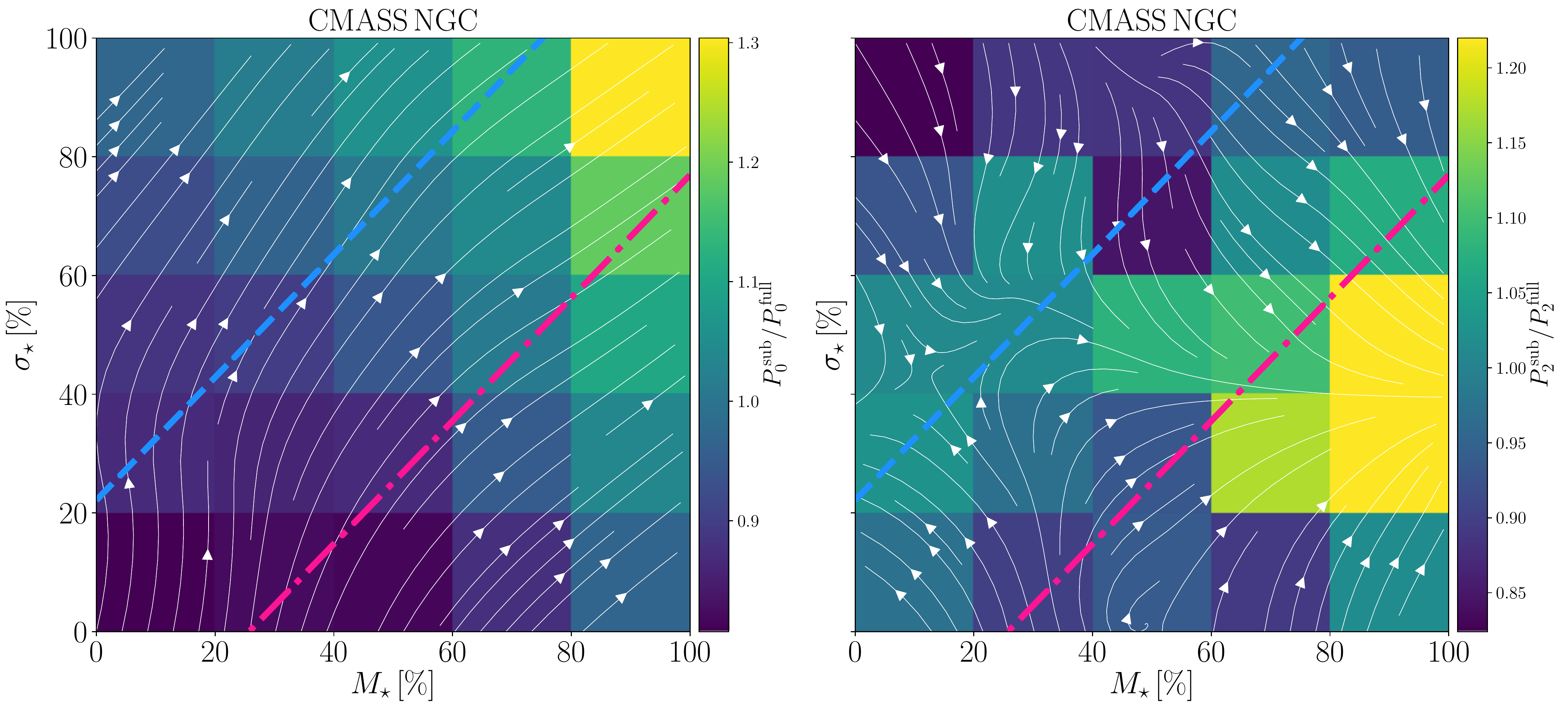}
\caption{Dependence of large-scale multipoles of the cross power spectrum on measured CMASS NGC galaxy properties -- stellar mass ($M_\star$) and stellar velocity dispersion ($\sigma_\star$). The linear extent of each pixel in this figure corresponds to a redshift-dependent quintile in each dimension ($M_\star$ and $\sigma_\star$) separately. The two panels show the mean monopole (left) and quadrupole (right) ratios of the subsamples with respect to the multipoles of the full sample. The color represents the mean amplitude with respect to the full sample on scales $k\leq0.2\,[h\,\mathrm{Mpc}^{-1}]$. Also shown are the gradients of the measured ratios (white lines with arrows). In both panels we show the two simple cuts we used to select objects with high $\sigma_\star$ and low $M_\star$ (above blue dashed line), and vice versa (below dot-dashed red line).}
\label{fig:Pixel}
\end{figure*}

\section{Splits based on galaxy velocity dispersion and stellar mass}  \label{sec:Matching_P0}

\subsection{Subsamples with matching redshift distributions}

We are interested in the dependence of the power spectrum multipoles on the galaxy properties, focusing in this section on velocity dispersion $\sigma_\star$ and stellar mass $M_\star$.  The most simple way to examine the dependence of clustering on these properties would be to split the galaxy samples into subsets based on $\sigma_\star$ and $M_\star$.  However, as discussed above (e.g., see Fig.\ \ref{fig:Mstar_sigmastar}), the distributions of these galaxy properties evolve significantly with redshift.  Therefore, straightforward cuts on these properties will produce samples with different redshift distributions, and since clustering evolves over redshift, it will be difficult to ascribe any difference in clustering to anisotropic bias rather than redshift evolution. In addition, each sample will have a different window function, making direct comparison difficult.

We therefore adopt a (slightly) more complicated approach to splitting samples, but that makes their analysis more simple.  We first bin our galaxies into $N_{\rm bin}=30$ redshift bins, spaced evenly across the relevant redshift range for each sample, $0.15<z<0.43$ for LOWZ and $0.43<z<0.7$ for CMASS.
Within each redshift bin, we rank-order the galaxies based on their properties, i.e.\ we convert their $\sigma_\star$ and $M_\star$ values separately into percentiles within each redshift bin.  We can then split the galaxies into subsets using their percentiles, rather than using fixed, redshift independent limits in $\sigma_\star$ and $M_\star$.  Creating subsamples this way ensures that they will always have redshift distributions matching the full sample (and matching each other). In order to avoid introducing binning effects, we interpolate the mapping between percentile and $\sigma_\star$ or $M_\star$ as a function of redshift, using linear interpolation between the discrete bin centers.  

To examine how the power spectra depend on $\sigma_\star$ and $M_\star$, we divide our sample into $5\times5$ pixels evenly divided along each dimension in the space of percentiles.  Note that this does not give 25 subsamples containing equal numbers of galaxies, since $\sigma_\star$ and $M_\star$ are correlated with each other.
For each percentile bin, we compute the multipoles of the cross power spectrum of galaxies belonging to that bin with the full sample. We then take the ratio with respect to multipoles of the full sample. To get an estimate of the amplitude of the multipoles, we compute the weighted average of the ratios of monopoles and quadrupoles at scales $k\leq0.2\,h\,\mathrm{Mpc}^{-1}$. Since we are computing both  power spectra at same scales, we weight the ratios by $k$ to account for the different number of modes in each $k$-bin. 

In Fig.\ \ref{fig:Pixel} we show the resulting dependence of the multipoles' amplitudes as a function of $M_\star$ and $\sigma_\star$ (in percentiles) for the CMASS NGC sample. Because all samples and sub-samples cover the same volume - their angular masks and redshift distributions match, so the window functions are the same and they can be directly compared. In the case of the monopoles, we find the expected dependence -- moving to larger values in both $M_\star$ and $\sigma_\star$ we obtain larger amplitudes. This arises because galaxies with larger stellar masses are expected to be hosted in more massive halos, thereby being more strongly biased, and vice versa. Similarly, more massive galaxies have larger velocity dispersion, so the monopole amplitude increases with $\sigma_\star$.  Note, however, that the quadrupole depends quite differently on $M_\star$ and $\sigma_\star$.

\begin{figure*}[!ht]
\includegraphics[width=0.96\textwidth]{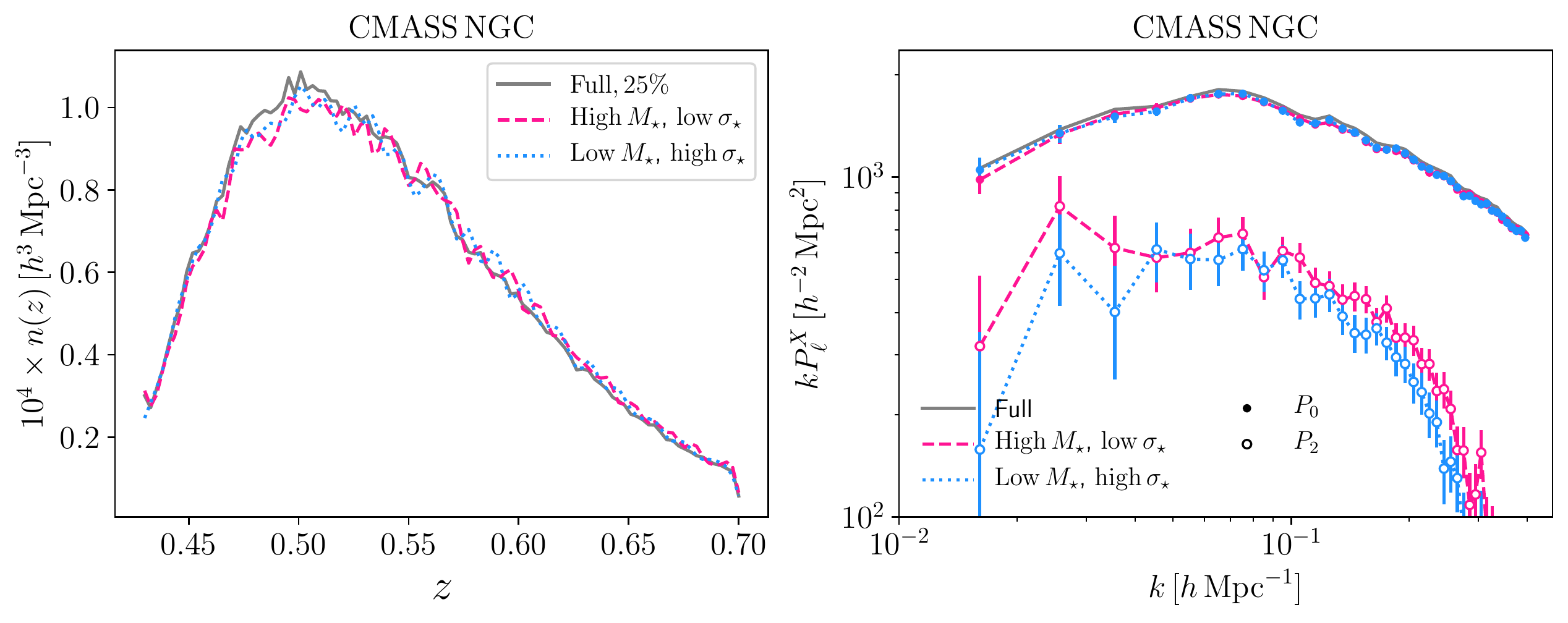}
\caption{\textit{Left panel:} Number density distribution as a function of redshift for the two CMASS NGC quartile subsamples. Solid gray line shows the 25\% scaled distribution of the full sample $n(z)$. \textit{Right panel:} Measured cross-power spectrum multipoles of the two quartile subsamples with the full sample, along with the full sample auto power spectrum monopole (solid gray line).}
\label{fig:nz_25_cmass}
\end{figure*}

\begin{figure*}[!ht]
\includegraphics[width=0.96\textwidth]{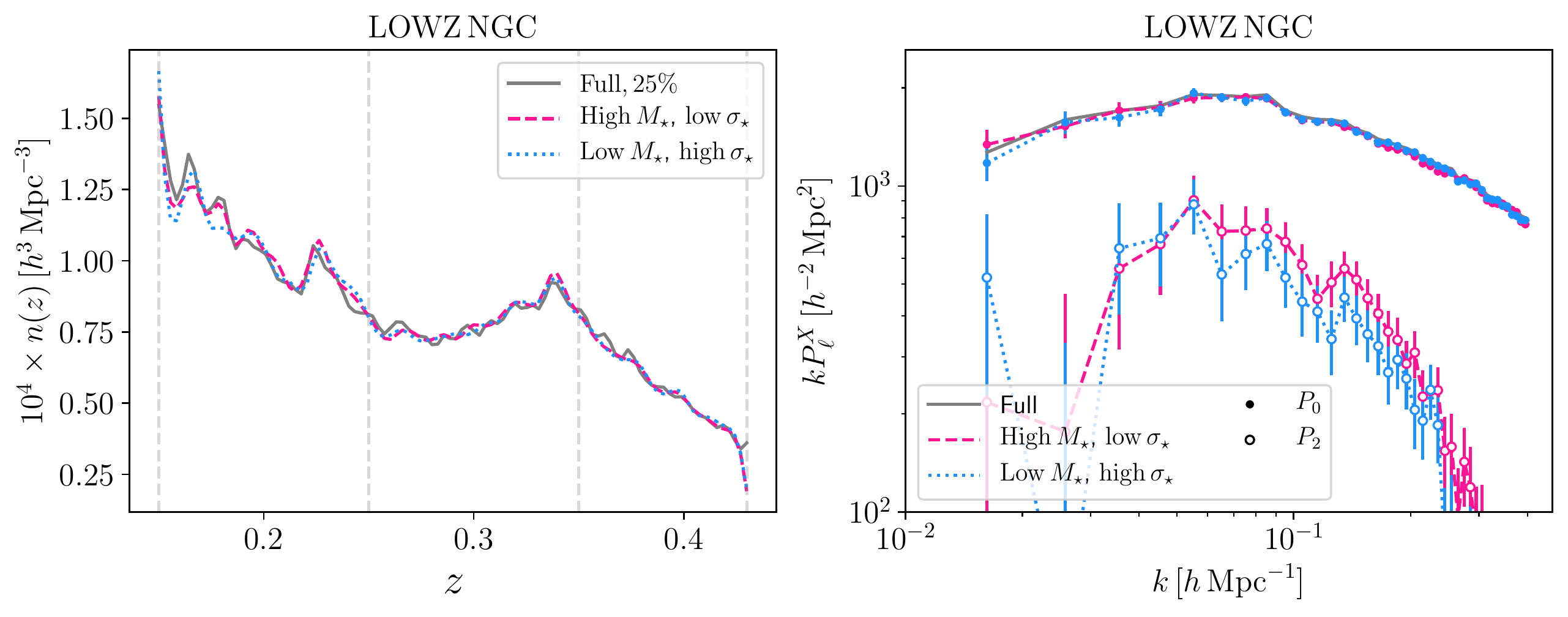}
\caption{\textit{Left panel:} Number density distribution as a function of redshift for the two LOWZ NGC quartile subsamples. Also shown is the 25\% scaled distribution of the full sample $n(z)$ (solid gray line). The vertical dashed lines are the redshift bins edges we used to perform the split. \textit{Right panel:} Measured cross-power spectrum multipoles of the two quartile subsamples with the full sample, along with the full sample auto power spectrum monopole (solid gray line).}
\label{fig:nz_25_lowz}
\end{figure*}

The significantly different dependence of the monopole and quadrupole on the galaxy properties $M_\star$ and $\sigma_\star$ is highly suggestive of AB.  As Fig.\ \ref{fig:Pixel} illustrates, the dependence of the quadrupole amplitude tends to be almost perpendicular compared to the one of the monopole. To make this more clear, we also show the 2D gradient of the multipoles in Fig.\ \ref{fig:Pixel} (white lines with arrows).  Although the gradients are noisy, especially for the quadrupole, we see that their directions are roughly perpendicular in the left and right panels.

\subsection{Subsamples with matching redshift distribution and monopoles}

Using the trends shown in Fig.\ \ref{fig:Pixel} as a guide, we construct two subsamples from the BOSS data based on galaxy properties $M_\star$ and $\sigma_\star$ with matching redshift distributions {\it and} matching monopoles. To accomplish this, we draw two straight lines in the percentile space of the monopole ratios (see Fig.\ \ref{fig:Pixel}) based on which we perform the sample cuts: one split corresponding to taking higher values of $M_\star$ and lower values of $\sigma_\star$, while the other has lower values of $M_\star$ and higher values of $\sigma_\star$ (see Fig. \ref{fig:Pixel}). We parametrize these cuts using: 
\be
\label{eq:line}
\sigma_\star\,[\%] = A\times M_\star\,[\%] + B,
\ee
where $A=\tan(\alpha)$ is the slope and $B$ is the intercept. We choose the slope to roughly match the slope of the monopole gradient and then we tune the intercept $B$ such that we obtain two disjoint subsamples with 25\% of galaxies in each. We provide the values of $\alpha$ and $B$ for both lines in Table \ref{table:1}.
Note that in designing these cuts, we use only the monopole measurements in Fig.\ \ref{fig:Pixel}, i.e.\ we do not use the quadrupole measurements. That is, our cuts are designed only to ensure that the two subsamples have consistent monopoles, and are not designed to separate the quadrupoles. 
Therefore, we can determine the significance of any difference in the quadrupoles without worrying about look-elsewhere effects from our choice of cuts.

While for CMASS we used redshift-independent cuts and obtained matching monopoles and $n(z)$, for LOWZ NGC we find that the trend between the amplitude of the monopole and $M_\star$ and $\sigma_\star$ varies significantly with redshift, such that a single redshift-independent cut does not cleanly split the sample. Therefore, we decompose the LOWZ sample into three redshift bins, measure the monopole ratios and perform the split in each bin. In Fig. \ref{fig:Pixel_LOWZN} we show the resulting dependence of the mulitpoles' amplitudes as a function of $M_\star$ and $\sigma_\star$ for the LOWZ NGC sample across three redshift bins. Finally, we merge the LOWZ subsamples across redshift bins to obtain the final LOWZ subsamples. 

In Fig.\ \ref{fig:nz_25_cmass} we show the galaxy redshift distributions and the measured cross power spectrum multipoles of the two subsamples with the full CMASS NGC sample. In Fig.\ \ref{fig:nz_25_lowz} we show the redshift distributions of galaxies and the measured cross power spectrum multipoles of the two subsamples with the full LOWZ sample.

Additionally, in order to test the impact of AC on the LOWZ sample, we repeat our analysis using $\sigma_\star^\mathrm{AC}$ (see Eqn.\ \eqref{eq:sigma_star_AC}). Lines that we used to perform this split differ from the ones we obtained when using $\sigma_\star$. We also provide these values in Table \ref{table:1}.

We note that we match the monopoles after a constant shot noise term has been subtracted (see Sec. \ref{sec:pk}). One possible concern could be that our two subsamples actually have very different shot noises. If such a difference exists, then the match between the shot-noise subtracted monopoles would not mean that they had the same bias and that they would be expected to have matching quadrupoles. This is not the case, however, as any wrongly estimated constant shot noise in the two subsamples would result in diverging monopoles on small scales, while we find that the monopoles of our subsamples agree on all scales for which we measure our power spectrum.

\begin{figure}[!ht]
\includegraphics[width=0.48\textwidth]{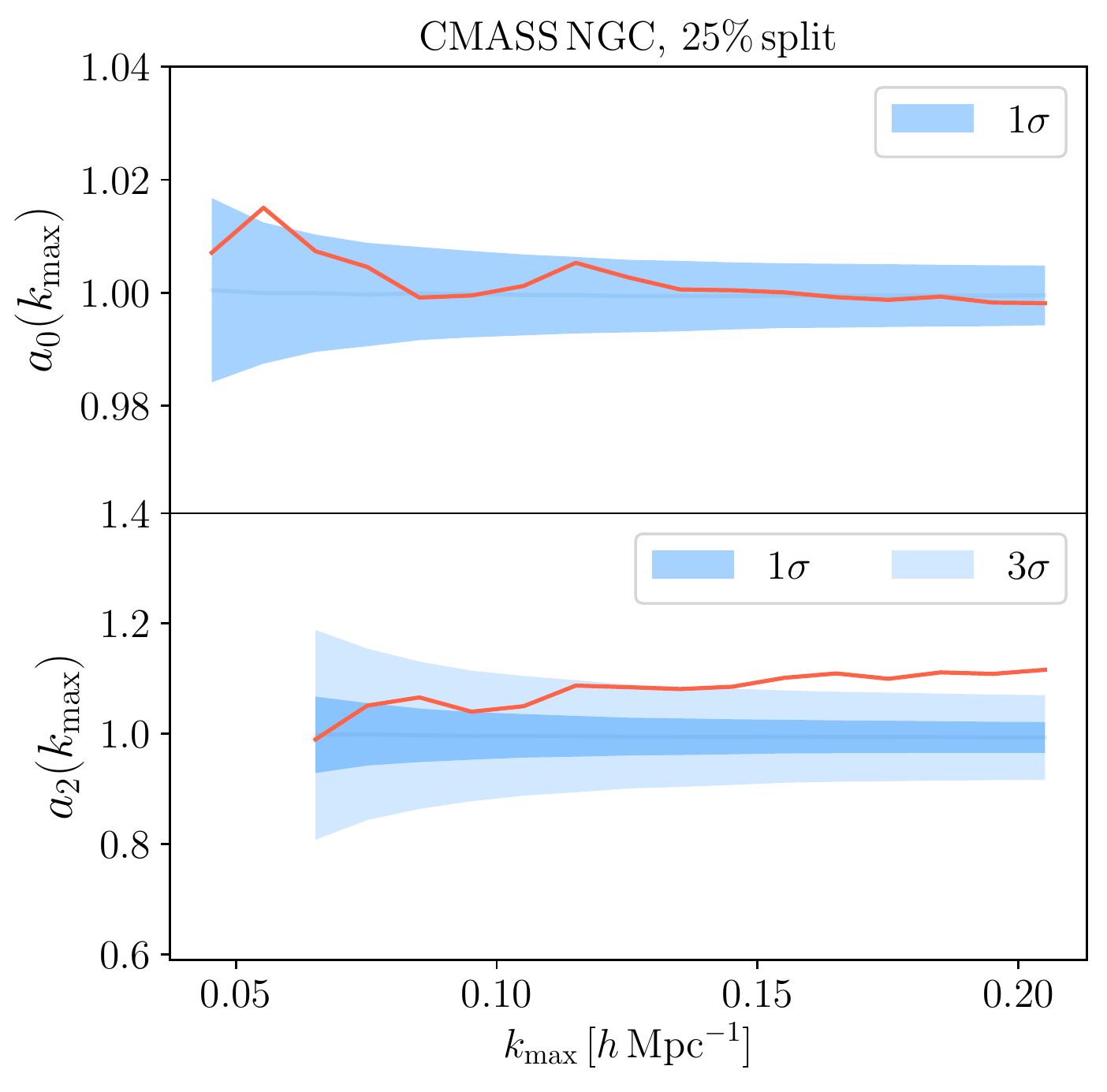}
\caption{Best fit parameter $a_\ell$ as a function of $k_\mathrm{max}$ using CMASS NGC. The top (bottom) panel shows $a_\ell(k_\mathrm{max})$ as fitted to the monopoles (quadrupoles). The shaded areas represent $1$ and $3\sigma$ regions obtained using mock catalogs and performing random splits matching the number density of the splits performed to the data.}
\label{fig:a_kmax_25_CMASSN}
\end{figure}

\begin{figure}[!ht]
\includegraphics[width=0.48\textwidth]{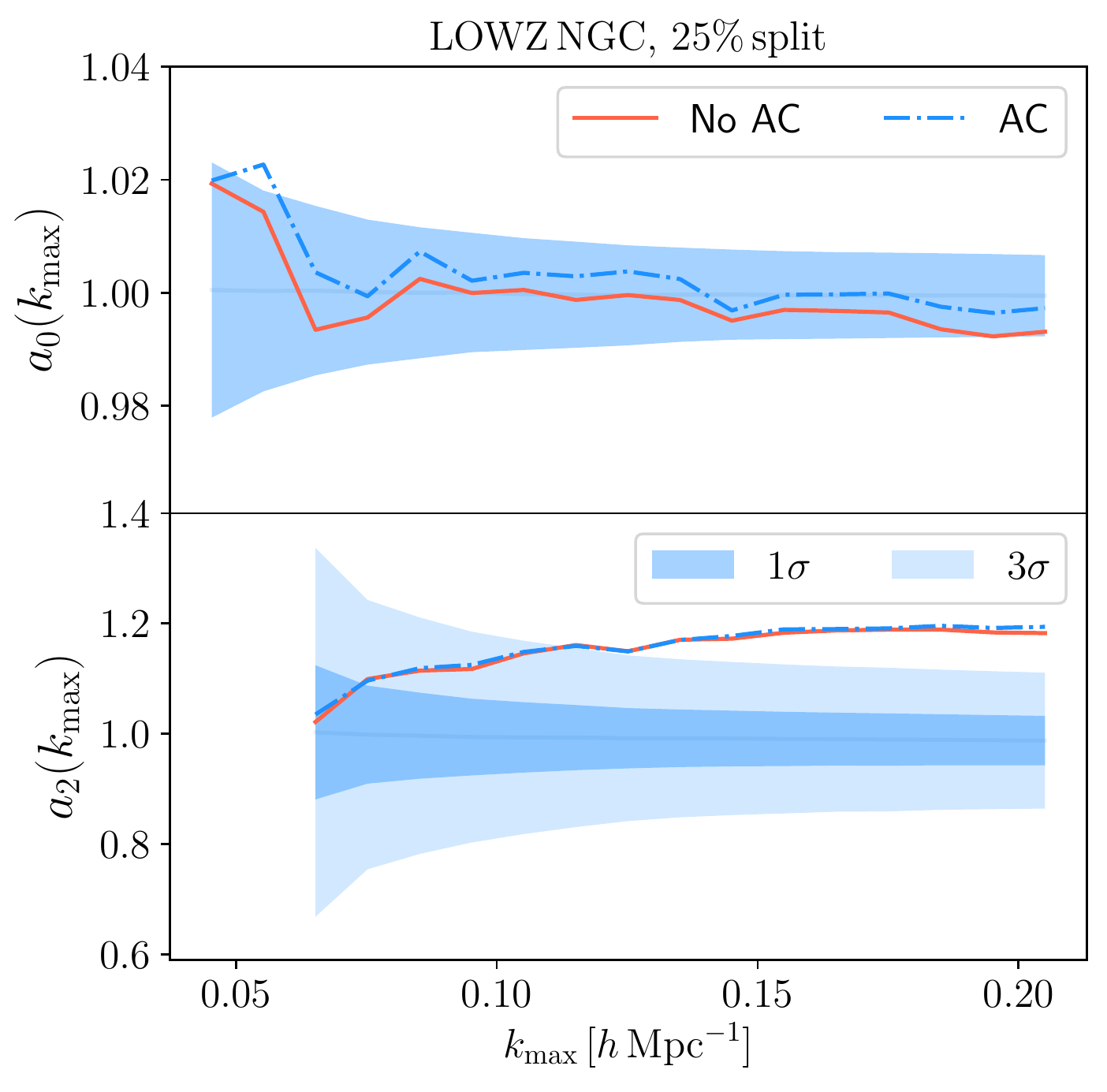}
\caption{Best fit parameter $a_\ell$ as a function of $k_\mathrm{max}$ using LOWZ NGC. The top (bottom) panel shows $a_\ell(k_\mathrm{max})$ as fitted to the monopoles (quadrupoles). The shaded areas represent $1$ and $3\sigma$ regions obtained using mock catalogs and performing random splits matching the number density of the splits performed to the data.}
\label{fig:a_kmax_25_LOWZN}
\end{figure}

\subsection{Detection significance}

Anisotropic bias of the form given in Eqn.\ \eqref{eq:delta}, with scale-independent bias parameters, will rescale the amplitude of the quadrupole without changing its shape.  We can therefore quantify the significance of AB by testing whether the measured power spectrum quadrupoles have different amplitudes, for two subsamples with matching monopoles and matching $n(z)$.  More precisely, we determine the value of the rescaling parameter $a_\ell$ which best brings the $\ell$ multipole of the two subsamples into agreement, i.e.\ we minimize:
\be \label{eq:chisq}
\chi^2(a_\ell)=[P_\ell^{\mathrm{sub},1}-a_\ell P_\ell^{\mathrm{sub},2}]^{\rm T} {\bf C}^{-1}_{a,\ell} [P_\ell^{\mathrm{sub},1}-a_\ell P_\ell^{\mathrm{sub},2}],
\ee
where ${\bf C}_{a,\ell}^{-1}$ is the inverse covariance matrix for the multipole $\ell$. We choose to measure the significance in this way, as we want to test the significance of the {\it amplitudes} being different, rather than the quadrupoles being different, which could have instead been tested using $\Delta P_\ell^{\rm T} {\bf C}^{-1}_{a,\ell}\Delta P_\ell$, where $\Delta P_\ell=P_\ell^{\mathrm{sub},1}- P_\ell^{\mathrm{sub},2}$. 

To estimate the covariance matrix ${\bf C}_{a,\ell}$ we use the available galaxy mocks described in \S\ref{mocks}, which do not include any AB.
We proceed as follows: Our null hypothesis is that there is no AB in the data, i.e.\ $b_q=0$ for both of the subsamples we analyse. That means that matching monopoles will have matching quadrupoles on large scales. Since we are keeping the monopoles fixed, we can test the significance of any detection of AB using samples created by randomly subsampling the mock catalogues. The distribution of results from these mocks gives the distribution from which our data measurement would be drawn if there were no AB. The level of significance at which the data disagrees with this distribution is the detection significance that we want to calculate.

Thus, in each mock we randomly select two disjoint subsamples with 25\% of the total galaxy sample to match in number of galaxies the data subsamples we analyse. We then measure the cross power spectrum multipoles of each subsample with the full mock. We repeat this for $N_m=1000$ mocks and construct the sample covariance matrix ${\bf C}_{a,\ell} = \langle \Delta P_{\ell}\, \Delta P_{\ell} \rangle$, with elements
\be
C_{a,\ell}^{ij} = \frac{1}{N_m-1}\sum_{m=1}^{N_m}  \Delta P_{m,\ell}(k_i)\ \Delta P_{m,\ell}(k_j),
\ee
where $\Delta P_{m,\ell}(k)=P^{\mathrm{sub},1}_{m,\ell}(k)- P^{\mathrm{sub},2}_{m,\ell}(k)$.  We expect that samples with matching multipoles should result in $a_\ell=1$ within the uncertainties.  We minimize $\chi^2(a_\ell)$ jointly fitting to both the monopoles and quadrupoles of both the samples. 

\begin{figure}[!ht]
\includegraphics[width=0.48\textwidth]{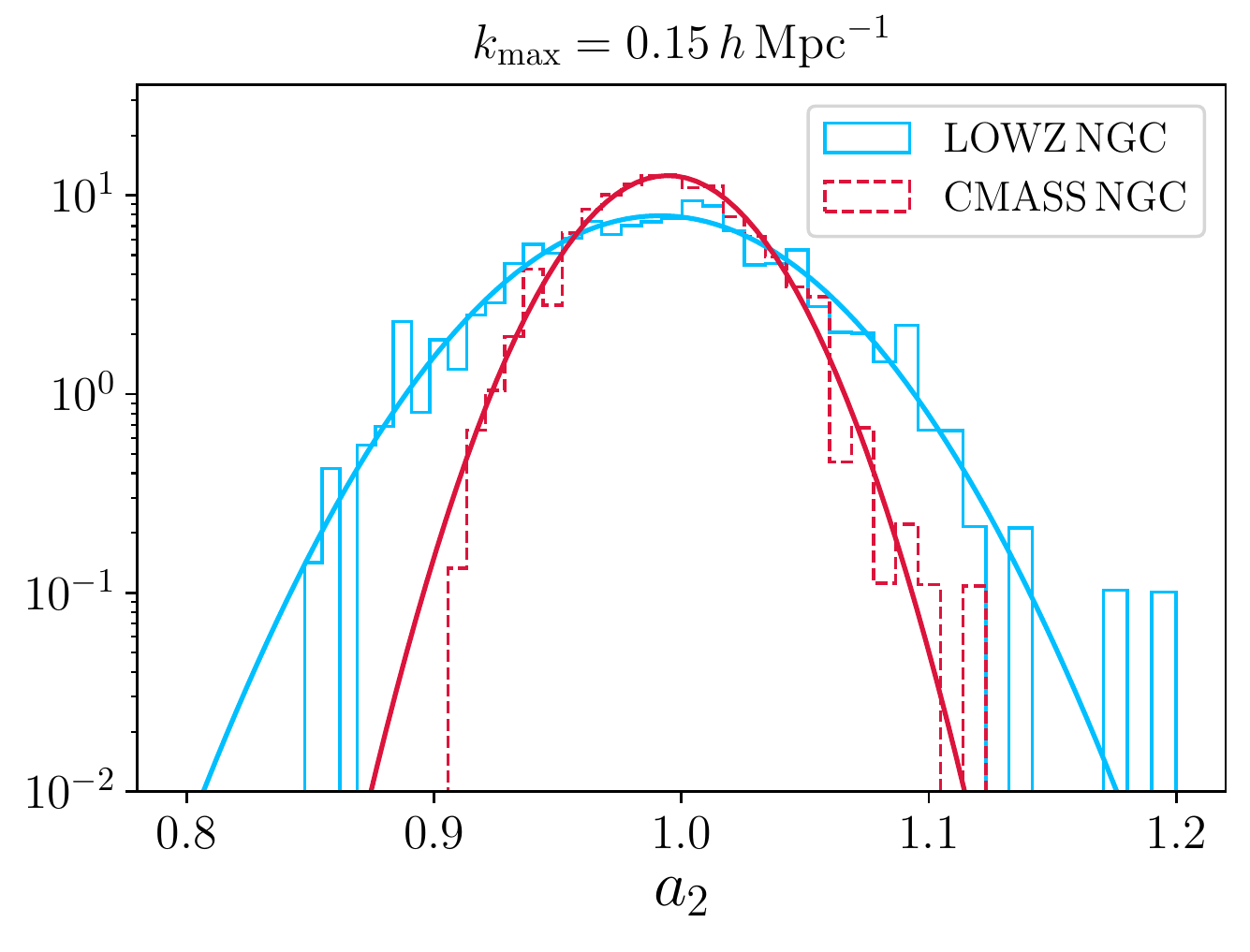}
\caption{Histogram of $a_2$ values obtained using mock catalogs using $k_\mathrm{max}=0.15\,h\mathrm{Mpc}^{-1}$ in the case of LOWZ NGC (blue) and CMASS NGC (red). For comparison we show the Gaussian probability distributions (solid lines) matching the mean and standard deviation of the measured $a_2$ distributions.} 
\label{fig:hist_a2}
\end{figure}

In Figures \ref{fig:a_kmax_25_CMASSN} and \ref{fig:a_kmax_25_LOWZN} we show the resulting values of $a_\ell$ as a function of the maximum $k$ fitted $k_\mathrm{max}$ in the case of CMASS and LOWZ NGC, respectively.  In order to determine whether these best-fitting values are consistent with no AB (i.e., $a_2=1$), we perform identical analyses on 1000 random mocks with $a_\ell=1$ and the same ${\bf C}_{a,\ell}$ as the real dataset, and compare the BOSS DR12 measurements to the mock results.  Because we only have $N_m=1000$ mocks, we cannot directly confirm confidence levels $\lesssim 10^{-3}$ by looking for numbers of inconsistent mocks.  However, we do find that the distribution of $a_l$ measurements from the mocks is quite consistent with a Gaussian distribution (see Fig. \ref{fig:hist_a2}).  Therefore, we assume that a measurement $|a_l-1|$ that deviates from 0 by more than $N$ times the rms from the mocks can be quoted as a detection significance of $N\sigma$.  In Figs.\ \ref{fig:a_kmax_25_CMASSN} and \ref{fig:a_kmax_25_LOWZN}, the shaded areas represent the $1$ and $3\,\sigma$ intervals of $a_\ell$ as a function of $k_\mathrm{max}$. For both CMASS and LOWZ, we find that our two subsamples are consistent with $a_0=1$ over a wide range of scales within $1\sigma$ uncertainty. On the contrary, the quadrupole difference results in best-fit values of parameter $a_2$ that are systematically different from one. For most of the $k_\mathrm{max}$ range, the inferred value of $a_2$ is $>3\sigma$ away from $a_2=1$.

\begin{figure}[!ht]
\includegraphics[width=0.48\textwidth]{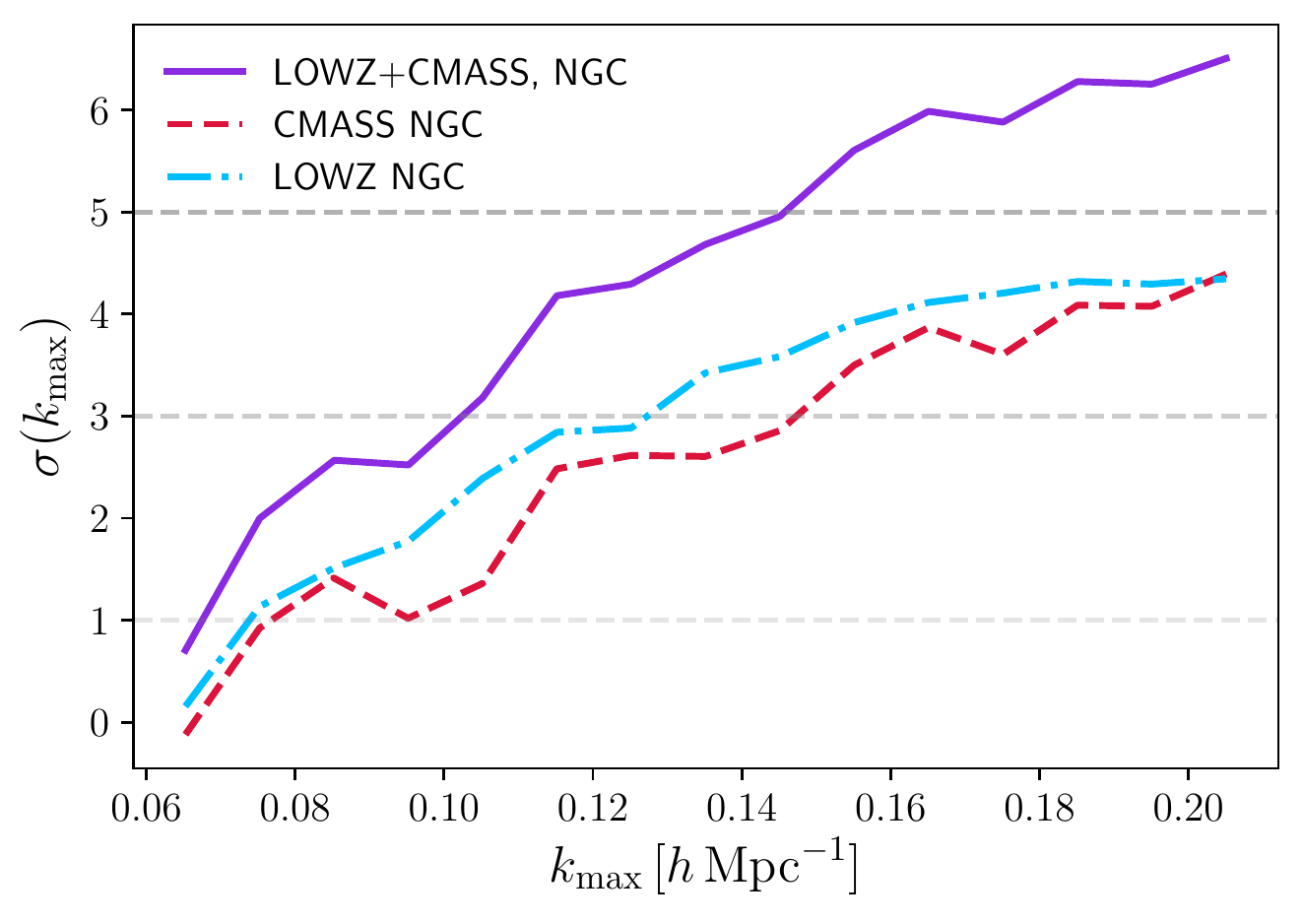}
\caption{Estimated significance of obtaining $a_2\ne1$ expressed in units of standard deviation, as a function of $k_\mathrm{max}$. Shown are the cases of using LOWZ and CMASS results individually (dot-dashed and dashed lines, respectively) and the combined significance (solid line).} 
\label{fig:Combined_akmax}
\end{figure}

We combine the results from the measurements of $a_2$ from LOWZ and CMASS to estimate the total significance. Since these galaxy samples are independent, we multiply the likelihoods to give the significance that we obtain a value equal to or larger than the measured $a_2$.  As noted above, we make no look-elsewhere corrections to the detection significance, since our cuts were designed without making reference to the quadrupoles. In Fig.\ \ref{fig:Combined_akmax} we show the significance of $a_2\ne1$ as a function of $k_\mathrm{max}$ in the cases of considering individual and combined measurements. We find the total significance is higher than $3\sigma$ when using $k_\mathrm{max}\approx0.11\,[h\,\mathrm{Mpc}^{-1}]$ and becomes $5\sigma$ when using $k_\mathrm{max}\approx0.15\,[h\,\mathrm{Mpc}^{-1}]$.

If we model the power spectrum multipoles using Eqn. \eqref{eq:P2D}, then the inferred anisotropic bias is approximately $\Delta b_q \equiv b_{q,1}-b_{q,2} \sim a_2-1$, for the observed values of $b_g$ and $f$.  
Note that we are only sensitive to the difference in anisotropic biases, $\Delta b_q$, and not to $b_{q,1}$ or $b_{q,2}$ separately, since for either sample alone, $b_q$ would be exactly degenerate with $b_g$ and $f$ on linear scales.  

\section{Discussion}  \label{sec:conc}

In this paper we have presented significant evidence for AB in BOSS DR12 galaxies, in both the LOWZ NGC and CMASS NGC samples. Our detection is the first to exceed the level of $5\sigma$, and was performed using a very simple test. By selecting subsamples from BOSS with the same redshift distribution and distribution of weights, we can analyse them robustly using the same pipeline. There is no difference in interpretation of power spectra as the window functions are the same. The test we perform is also very simple - without AB, samples with the same monopole have the same large-scale bias and $f$. If the large-scale quadrupole only depends on these quantities, as is the case in the absence of AB, we would expect consistent quadrupoles. Instead, we find inconsistent quadrupoles using BOSS samples split in $\sigma_\star$ and $M_\star$.  This result is qualitatively consistent with the behavior of dark matter halos in N-body simulations, although the magnitude of the AB we detect in BOSS galaxies is far smaller than the magnitude seen in simulated halos \cite{Obuljen}, suggesting large misalignments between the motions of stars in massive elliptical galaxies and the motions of dark matter particles in their host halos.

An obvious follow-up question is whether we can detect AB using other properties besides velocity dispersion.  We discuss this in more detail in Appendix \ref{sec:Matching_P2}, where we consider additional properties including projected galaxy size $R_0$ and surface brightness $I_0$. We see a strong trend in the monopole with $R_0$, which has a higher amplitude for samples selected with larger $R_0$, as expected for larger galaxies. However, there is no clear trend for the quadrupole unlike for samples split using $\sigma_\star$, suggesting that $R_0$ is not as tightly coupled to the tidal field as the velocity dispersion, leading to smaller AB effects.

Extending this analysis to samples split using a combination of $I_0$, $R_0$ and $\sigma_\star$, based on the Fundamental Plane, we find that the results are very sensitive to the parameters assumed for the Fundamental Plane and, depending on the combination of parameters chosen, results range from splits showing large amounts of AB, or no significant AB.  This wide range of behavior perhaps might explain why previous searches for AB using the Fundamental Plane obtained results inconsistent with each other \cite{Hirata_obs,Singh}, since different choices for the tilt of the Fundamental Plane produce quite different levels of AB. These results are consistent with our primary analysis which split the sample using $\sigma_\star$: different selection methods lead to different levels of correlation between the galaxies in the sample and the large-scale tidal fields, and different levels of completeness in sampling galaxies with all orientations. 

As we noted above, our methodology is sensitive only to differences in AB between two samples, $\Delta b_q$, and not the absolute level of AB in either sample. For the BOSS DR12 sample that we have analyzed, we found $\Delta b_q\approx 0.1-0.2$ between subsets with high $M_\star$ and low $\sigma_\star$, and subsets with low $M_\star$ and high $\sigma_\star$. For comparison, the overall linear bias for this galaxy population is $b_g \approx 2$.  As is evident from Fig.\ \ref{fig:Pixel}, we could have chosen a different split to give two subsets with $\Delta b_q \approx 0$, e.g.\ by choosing cuts oriented perpendicularly to our cuts.  This would not necessarily mean that AB is absent in those subsets, only that there is no detectable difference in AB between them.  Our measurements therefore cannot be used to place upper limits on the magnitude of AB $|b_q|$ for any sample.  Instead, one could use N-body simulations as a guide for deriving priors on the size of $b_q$ for dark matter halos, under the assumption that galaxies can only have smaller large-scale $b_q$ than their host halos. 

The results from the splits do give a way to estimate the potential level of contamination for selections that depends strongly on $\sigma_\star$. We find that, for halos in the mass range relevant for LRGs, $|b_q|$ can exceed the growth rate $f$ \cite{Obuljen} depending on the halo selection.  The only way to be certain that a population has no AB is to ensure that the population is complete, i.e.\ it is selected using only intrinsic scalar properties (like mass) and is independent of orientation, shape, motion, etc.\ \cite{Hirata_th}. Since AB is degenerate with redshift-space distortions, this implies that great care must be taken when interpreting RSD measurements of real, observed galaxies. In addition, when designing a sample to be observed one should take care to only select on scalar properties that are independent of orientation.

\begin{acknowledgments}
We thank Faizan Mohammad, Marko Simonovi\'{c}, Chris Duckworth \& Rita Tojeiro for useful discussions. This research was supported by the Centre for the Universe at Perimeter Institute. Research at Perimeter Institute is supported in part by the Government of Canada through the Department of Innovation, Science and Economic Development Canada and by the Province of Ontario through the Ministry of Colleges and Universities. We acknowledge support provided by Compute Ontario (www.computeontario.ca) and Compute Canada (www.computecanada.ca). We also acknowledge the use of \texttt{nbodykit} \cite{nbodykit}, \texttt{IPython} \cite{IPython}, \texttt{Matplotlib} \cite{Matplotlib}, \texttt{NumPy} \cite{Numpy} and \texttt{SciPy} \cite{SciPy}.

Funding for SDSS-III has been provided by the Alfred P. Sloan Foundation, the Participating Institutions, the National Science Foundation, and the U.S. Department of Energy Office of Science. The SDSS-III web site is http://www.sdss3.org/.

SDSS-III is managed by the Astrophysical Research Consortium for the Participating Institutions of the SDSS-III Collaboration including the University of Arizona, the Brazilian Participation Group, Brookhaven National Laboratory, Carnegie Mellon University, University of Florida, the French Participation Group, the German Participation Group, Harvard University, the Instituto de Astrofisica de Canarias, the Michigan State/Notre Dame/JINA Participation Group, Johns Hopkins University, Lawrence Berkeley National Laboratory, Max Planck Institute for Astrophysics, Max Planck Institute for Extraterrestrial Physics, New Mexico State University, New York University, Ohio State University, Pennsylvania State University, University of Portsmouth, Princeton University, the Spanish Participation Group, University of Tokyo, University of Utah, Vanderbilt University, University of Virginia, University of Washington, and Yale University.

The massive production of all MultiDark-Patchy mocks for the BOSS Final Data Release has been performed at the BSC Marenostrum supercomputer, the Hydra cluster at the Instituto de Fısica Teorica UAM/CSIC, and NERSC at the Lawrence Berkeley National Laboratory. We acknowledge support from the Spanish MICINNs Consolider-Ingenio 2010 Programme under grant MultiDark CSD2009-00064, MINECO Centro de Excelencia Severo Ochoa Programme under grant SEV- 2012-0249, and grant AYA2014-60641-C2-1-P. The MultiDark-Patchy mocks was an effort led from the IFT UAM-CSIC by F. Prada’s group (C.-H. Chuang, S. Rodriguez-Torres and C. Scoccola) in collaboration with C. Zhao (Tsinghua U.), F.-S. Kitaura (AIP), A. Klypin (NMSU), G. Yepes (UAM), and the BOSS galaxy clustering working group.
\end{acknowledgments}

\appendix

\section{Sample cuts}
\label{sec:Cuts}
In \S\ref{sec:Matching_P0} we used straight cuts to match the monopoles and quadrupoles of the subsamples, respectively. Here, in Table \ref{table:1}, we provide the values of $\alpha$ and $B$ we used in Eqn.\ \eqref{eq:line} to perform these cuts. We denote with $B^+$ and $B^-$ the intercept values used when obtaining the sample with respectively higher and lower values of $\sigma_\star$.

\begin{table}[!ht]
\label{table:1}
  \centering
  \begin{tabular}{|c|c|c|c|}
  \hline
$\sigma_\star -M_\star$ & $z_\mathrm{bin}$ & $\alpha[^{\circ}]$ & $B^+ / B^-\,[\%]$\\
\hline
CMASS & 0.43 -- 0.70 & 46 & 22.15 / -26.63 \\ 
\hline
\multirow{3}{*}{LOWZ, no AC} & 0.15 -- 0.25 & \multirow{3}{*}{45} & 19.04 / -19.33 \\ \cline{2-2} \cline{4-4}
& 0.25 -- 0.35 &  & 21.38 / -21.09 \\ \cline{2-2} \cline{4-4}
& 0.35 -- 0.43 &  & 23.72 / -23.72 \\
\hline
\multirow{3}{*}{LOWZ, AC} & 0.15 -- 0.25 & \multirow{3}{*}{46} & 20.74 / -18.00 \\ \cline{2-2} \cline{4-4}
& 0.25 -- 0.35 &  & 23.18 / -19.52 \\ \cline{2-2} \cline{4-4}
& 0.35 -- 0.43 &  & 25.32 / -21.66 \\
\hline
\end{tabular}
\caption{Line parameters used to perform $\sigma_\star-M_\star$ splits. The first column shows the galaxy sample and values used; second column shows the redshift range in which the cut was performed; third column contains the slope parameter of the lines used; the last column contains the used intercept values.}
\end{table}

\begin{figure*}[!ht]
\includegraphics[width=0.96\textwidth]{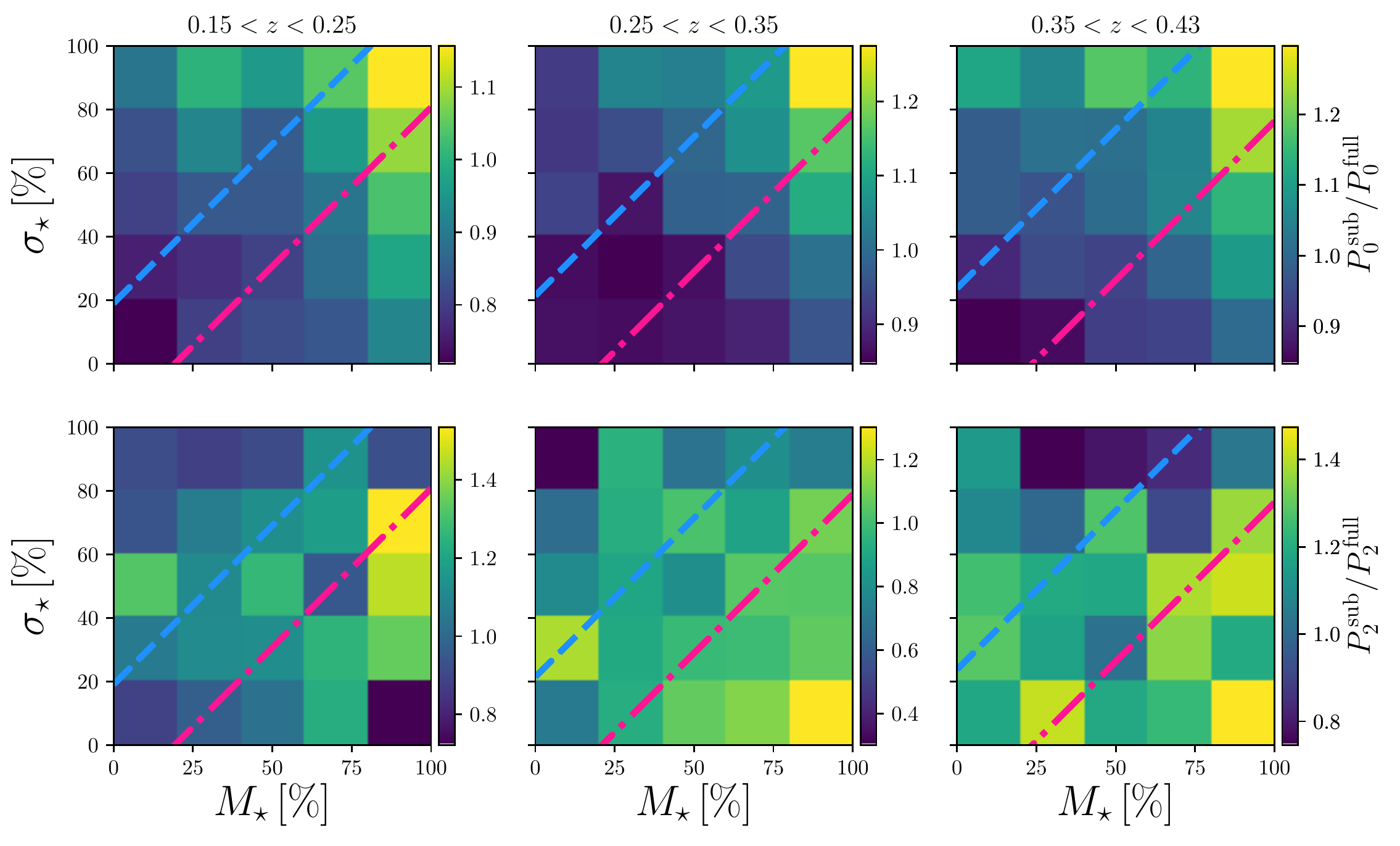}
\caption{Similar to Fig. \ref{fig:Pixel}, in the case of LOWZ NGC using three redshift bins. Top and bottom panels show, respectively, the monopole and quadrupole ratios of the subsamples with respect to the multipoles of the full sample. Panels from left to right correspond to different redshift bins.}
\label{fig:Pixel_LOWZN}
\end{figure*}

\section{Splits using other galaxy properties}
\label{sec:Matching_P2}
In \S\ref{sec:Matching_P0}, we presented evidence for AB, by splitting galaxies based on their stellar masses $M_\star$ and their line-of-sight velocity dispersions $\sigma_\star$.  Besides these properties, the BOSS catalog lists other properties as well, so it is worthwhile to explore whether similar signals of AB may be detected using any of those other properties.

\begin{figure*}[!ht]
\includegraphics[width=0.94\textwidth]{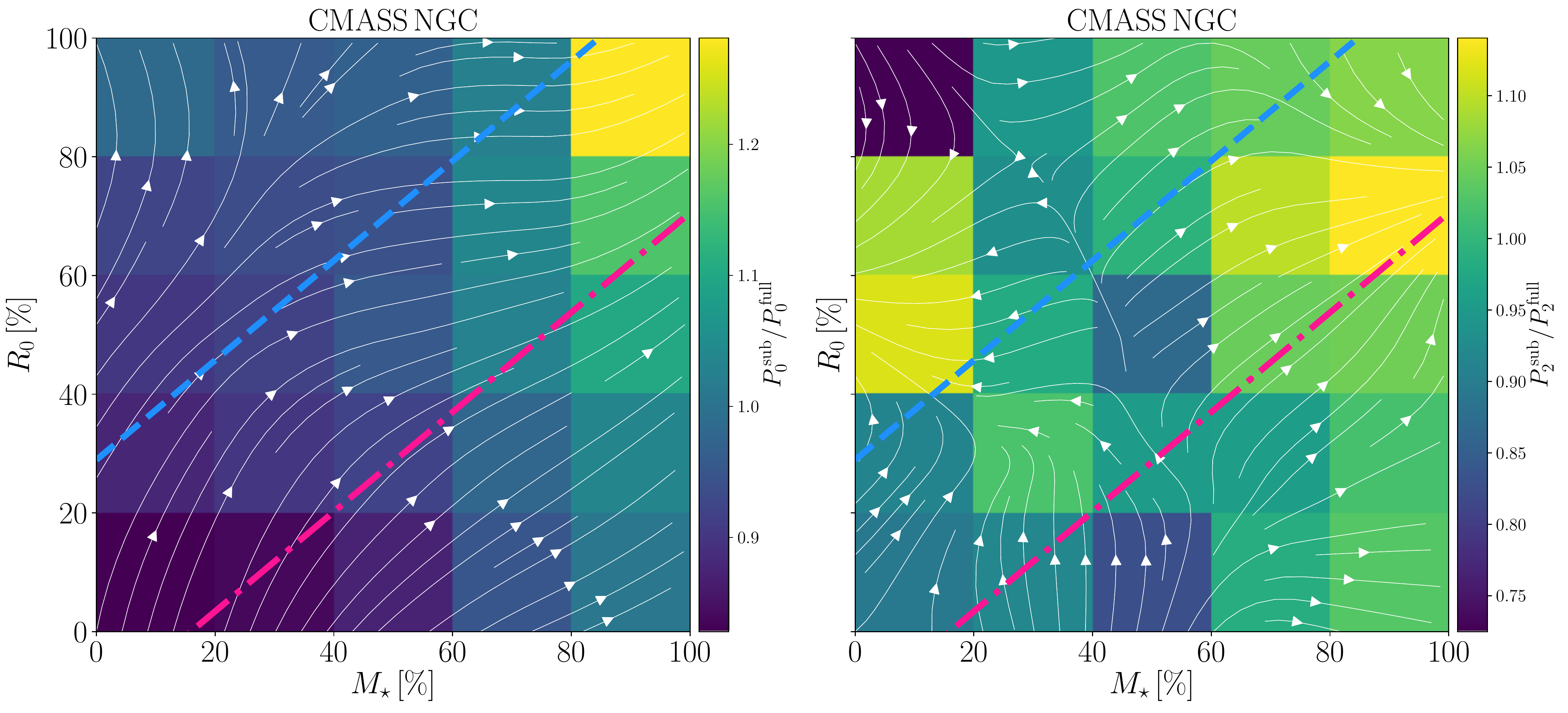}
\caption{Similar to Fig. \ref{fig:Pixel}, but now using the galaxy physical size ($R_0$) and stellar mass ($M_\star$).}
\label{fig:Pixel_R0_Ms}
\end{figure*}

\subsection{Splitting using the projected size}

As a first example, we can consider splitting galaxies using  $M_\star$ and $R_0$, the projected size of their stellar population.  In Fig.\  \ref{fig:Pixel_R0_Ms} we show the dependence of clustering amplitude as a function of $R_0$ and $M_\star$ (in percentiles) for the CMASS NGC sample, using the same procedure used to generate Fig.\ \ref{fig:Pixel}.
We see that the monopole behaves in the expected way, increasing in amplitude with increasing $M_\star$ and increasing $R_0$.  For the quadrupole, however, there is no significant trend seen in this space of $M_\star$ and $R_0$, quite different to what we found in Fig.\ \ref{fig:Pixel} using $M_\star$ and $\sigma_\star$.  The different dependence of the quadrupole on $R_0$ and $\sigma_\star$ is somewhat surprising, because in N-body simulations halo shapes are more strongly correlated with tidal fields than halo velocity dispersions are, e.g.\ Fig.\ 1 of Ref.\ \cite{Obuljen}.  The lack of any strong trend with $R_0$ in BOSS galaxies may suggest that elliptical galaxy shapes correlate more weakly with large-scale tides than galaxy velocity dispersions do, in contrast to the behavior expected for their host dark matter halos.  One possible reason for this may be that in both CMASS and LOWZ, the fractional scatter in $R_0$ is significantly larger than the corresponding scatter in $\sigma_\star$, which could act to wash out any correlations with large-scale tidal fields. Indeed, we find the fractional scatter to be smaller in $\sigma_\star$ compared to $R_0$ measurements: $\sigma(\log_{10}\sigma_\star)=0.084(0.151)$ and $\sigma(\log_{10}R_0)=0.197(0.212)$ for LOWZ (CMASS) NGC samples.

Because we find no strong dependence of the quadrupole in Fig.\ \ref{fig:Pixel_R0_Ms}, this means that essentially any cut that bisects the sample will give two subsets with matching quadrupoles.  Our test for anisotropic bias, in which we construct two subsets with matching monopoles and different quadrupoles, will therefore necessarily give a null result.  We have verified this by repeating the procedure from the previous section, and as expected we find no evidence for AB using $R_0$ and $M_\star$. 

More generally, we can also search for AB using not only pairs of galaxy properties, but other combinations.  For example, previous works \cite{Hirata_obs,Singh} have attempted to detect AB using the Fundamental Plane, a combination of 3 properties: $R_0$, $\sigma_\star$, and $I_0$, the projected surface brightness. These previous analyses found results somewhat in tension with each other, with \cite{Hirata_obs} reporting marginal ($2.3\sigma$) evidence for AB, whereas \cite{Singh} found no significant evidence for AB.  In order to address these, we perform the following analysis using both the LOWZ and CMASS NGC samples.

\subsection{Splitting using the Fundamental Plane}
\label{sec:FP}

\begin{figure}[!ht]
\subfloat{\includegraphics[width=0.48\textwidth]{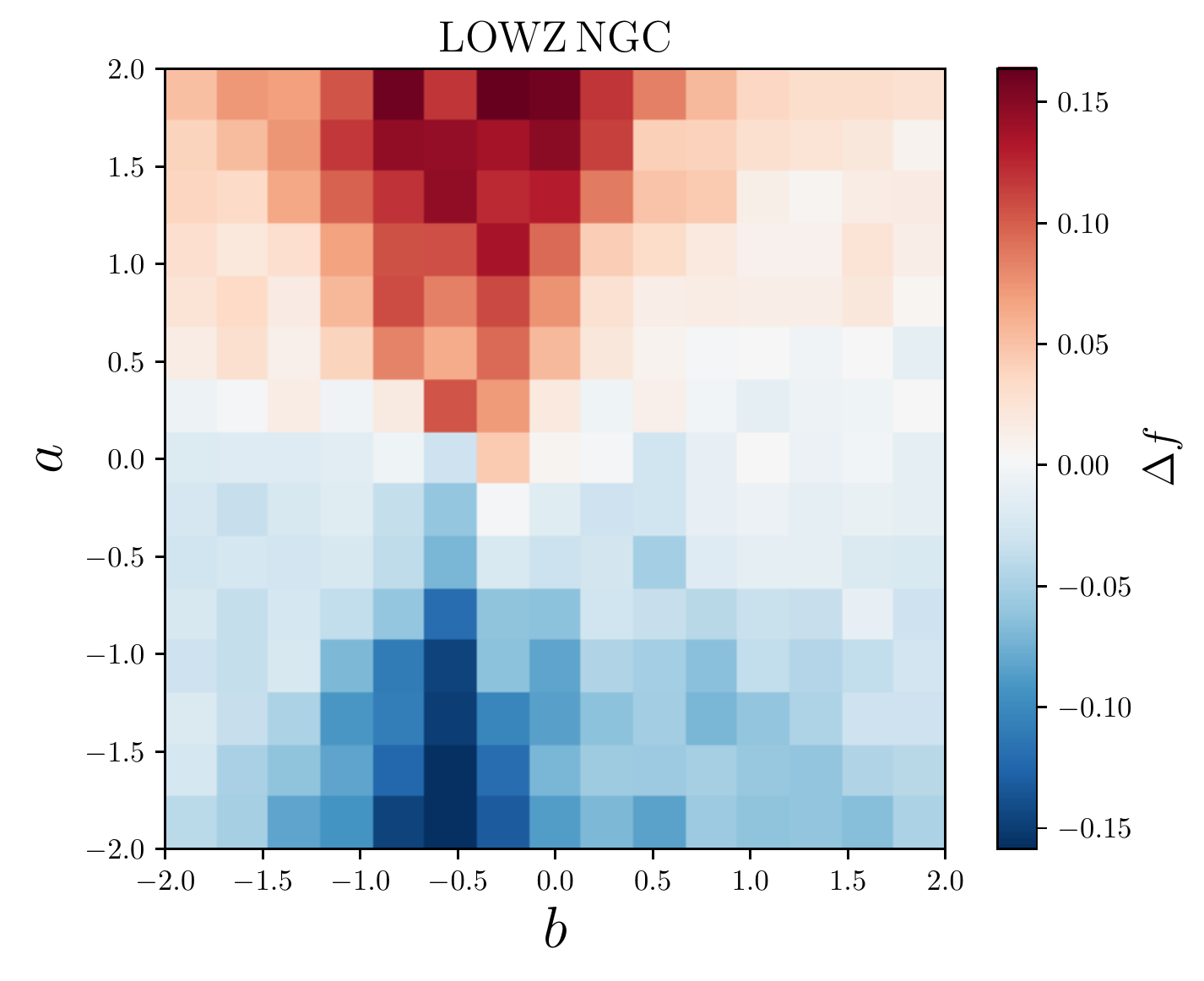}}\\
\subfloat{\includegraphics[width=0.48\textwidth]{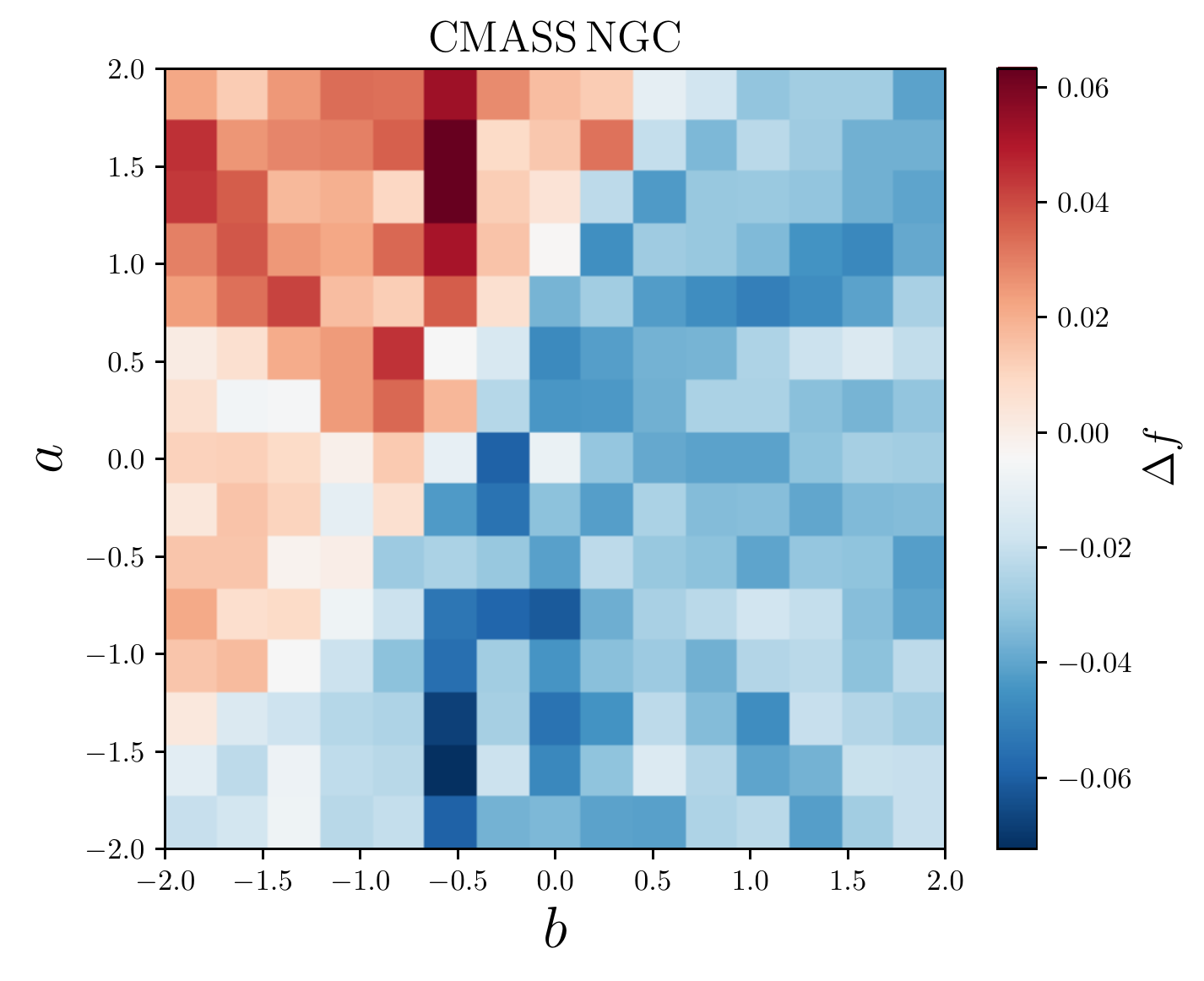}}
\caption{Differences in the inferred growth rate parameter $f$ between two samples obtained by splitting the full LOWZ NGC (top panel) and CMASS NGC (bottom panel) sample in half based on the FP residuals.}
\label{fig:FP}
\end{figure}

The Fundamental Plane (FP) is a relation between the galaxy surface brightness $I_0$, velocity dispersion $\sigma_\star$ and physical radius $R_0$, defined as:
\be
\log_{10}R_0=a\log_{10}\sigma_\star + b\log_{10}I_0 + c,
\ee
where $a,b,c$ are the FP parameters, chosen to minimize the scatter perpendicular to the FP. These parameters are expected to vary as a function of redshift, and possibly as a function of galaxy sample.  For the velocity dispersion we use the aperture corrected $\sigma_\star^\mathrm{AC}$ (see Eqn.\ \eqref{eq:sigma_star_AC}). 
We compute the surface brightness $I_0$ as:
\be
\log_{10}I_0 = -M_{ke}/2.5 - 2\log_{10}R_0 + 4\log_{10}(1+z),
\ee
where $M_{ke}$ is the $k+e$ corrected (at $z=0.55$) absolute magnitude in $i$-band that we obtain from the catalog of the Granada group\footnote{\href{https://www.sdss.org/dr12/spectro/galaxy_granada/}{https://www.sdss.org/dr12/spectro/galaxy\_granada/}} \cite{Granada}. 

For each galaxy, in either LOWZ NGC or CMASS NGC sample, we compute the ratio $I_0^b\sigma_\star^a/R_0$ based on which we split the full sample into two subsamples --- bottom and top 50\%. To account for redshift dependence of the FP, we perform this split in $20$ narrow redshift bins. We repeat this for a range of values of FP parameters $a$ and $b$. For each combination we measure the auto power spectrum multipoles of the two subsamples. We cannot adjust the samples to give consistent monopoles, and so in order to test for AB, we need to fit models to the data. We do this for models excluding AB, and measure the difference between recovered values of the growth rate $f$ from the subsamples.

To model the monopole and quadrupole measurements of each subsample's auto power spectrum we use linear theory \cite{Kaiser}:
\be
\begin{split}
P_0(k) &=\left(b_g^2+\frac{2}{3} f b_g + \frac{1}{5}f^2\right)P_m(k),\\
P_2(k) &=\left(\frac{4}{3} b_g f+\frac{4}{7}f^2\right)P_m(k).
\end{split}
\ee
Since this model is only valid on largest scales, we only use the scales in the range $0.04<k[\,h\,\mathrm{Mpc}^{-1}]<0.1$. We emphasize that our goal is not to obtain the true values of $f$, as is to obtain an estimate of the difference $\Delta f$ between the two subsamples after the FP split.

To estimate the covariance matrix for the FP splits, we cannot rely on the galaxy mocks, as we did in previous sections, where we randomly subsampled mock catalogs to match the monopole's amplitude between the two subsamples. Instead, we use an analytical Gaussian covariance matrix \cite{Grieb}, using the implementation in Ref. \cite{pyrsd}. This approach takes into account the survey varying $n(z)$ and is model dependent.

Our goal is to minimize $\chi^2(\bm{\theta})=\Delta P_\ell^T {\bf C}_\ell^{-1}(\bm{\theta}) \Delta P_\ell$, where $\Delta P_\ell$ is a vector containing the difference between the data and the model for both the monopole and the quadrupole, while our free parameters vector is $\bm{\theta} = \{b_g,f\}$. To minimize $\chi^2$, we use the Nealder-Mead method \cite{nelder1965simplex} as implemented in \texttt{SciPy} \texttt{minimize} function \cite{SciPy}. We repeat this procedure for various values of FP parameters $(a,b)$ and obtain the best-fit $\bm{\theta}$ for each subsample. 

In Fig. \ref{fig:FP} we show the resulting difference of growth rates from the two subsamples after the FP split, obtained using either LOWZ NGC or CMASS NGC. We find that for some values of $a,b$ the resulting $\Delta f$ is indeed consistent with non-detection of AB. However, there seems to be a range of $a,b$ values, for which an FP split would result in a rather significant detection of $\Delta f$.  Therefore, it appears that the choice of FP parameters strongly affects whether the FP can be used or not to detect AB.

\bibliography{References.bib}

\end{document}